\documentclass{aastex631} 
\usepackage{graphicx}
\usepackage{wrapfig}
\usepackage{txfonts}
\usepackage{ulem}
\usepackage{epstopdf}
\usepackage{multirow}
\usepackage{booktabs}
\usepackage{xcolor,colortbl}
\usepackage{rotating}
\usepackage[english]{babel}
\usepackage{todonotes}
\usepackage{natbib,twoopt}
\usepackage{textgreek}

\usepackage[normalem]{ulem}

\received{2020 July 28}
\revised{2021 January 21}
\accepted{2021 January 29}
\published{2021 March 16}

\shorttitle{Astrochemical Bistability}
\shortauthors{Dufour \& Charnley}

\begin{document}

\title{New Bistable Solutions in Molecular Cloud Chemistry: Nitrogen and  Carbon Autocatalysis } 
\author{Gwena\"elle  Dufour}
\affiliation{Astrochemistry Laboratory, Code 691, NASA Goddard Space Flight Center, Greenbelt, MD 20771, USA }
\affiliation{Department of Physics, Catholic University of America, Washington, DC 20064, USA}
\affiliation{Leiden Observatory, Leiden University, Leiden, The Netherlands}

\author{Steven B. Charnley}
\affiliation{Astrochemistry Laboratory, Code 691, NASA Goddard Space Flight Center, Greenbelt, MD 20771, USA }


\begin{abstract} 
We have investigated the chemistry of dense interstellar clouds and found new bistable solutions in the  nitrogen and carbon chemistries. 
We identify the autocatalytic processes that  are present in the pure, reduced, chemical networks and, as previously found for oxygen chemistry,  that He$^+$ plays an important role.  
 The applicability of these results to astronomical environments is briefly discussed. The bistable solutions found for carbon chemistry occur for low densities and  high ionization fractions that  are not compatible with that found cold, dense clouds.  Bistability in the pure nitrogen chemistry occurs for conditions that are relevant for prestellar cores in which significant CO depletion has taken place. 
We conclude that several autocatalyses are embedded in gas-phase interstellar chemistry and that many more are potentially present. 
\end{abstract}


\section{Introduction}

 Nonlinear chemical systems far from thermodynamic equilibrium, such as those commonly found in in astrophysical environments,  can in principle exhibit multistability and complex dynamical evolution, including bistability, oscillations and chaos   (e.g. \citet{1996ApJ...100..31E}). 
A large number of chemical models of dense interstellar clouds have been developed since early 1970s (\citet{2013ChRv..113.8710A}) and astrochemical bistability was first detected as steady-state  solutions to their gas-phase kinetics almost 30 years ago  (\citet{1992MNRAS.258P..45P,1993ApJ...416L..87L}).  
When a bifurcation occurs as a control parameter is varied, a nonlinear system may switch from a single stable solution to multiple simultaneous possible solutions, among which two at least are stable. This leads to a transition from one stable branch of a bifurcation diagram to another stable branch, connected by an unstable one.
The bistable solution discovered by  \citet{1993ApJ...416L..87L} was considered in several subsequent studies that explored the the appearance/disappearance of the bifurcation subject to variation of  physical and chemical control parameters relevant for dense molecular clouds (\citet{1995A&A...296..779S,1995A&A...297..251L, 1995A&A...302..870L,1998A&A...334.1047L, 2000RSPTA.358.2549P,2003A&A...399..583C,2006A&A...459..813W}). 
  
In  \citet{2019ApJ...887...67D} (Paper I) we proposed an explanation of dense cloud bistability.   
By deconstructing a known bistable solution  in a dense cloud chemical model  into ever simpler {\it reduced} models , through omission and/or depletion of chemical elements, followed by removing selected reactions in {\it artificial} models, we  identified autocatalytic processes involving oxygen nuclei,  related the formation efficiency of oxygen dimer (O$_2$) through chemical feedback, as the cause of this  bistability. This was demonstrated both in a reduced model containing only oxygen chemistry and in the original  dense cloud model.  
We showed that autocatalysis could  account for the known dependence on various control parameters  -   the ratio of the cosmic-ray ionization rate to the cloud number density ($\zeta/n_ {\rm H}$),  the He  ionization rate, the relative elemental depletions of carbon, oxygen and sulfur, and  the H$_3^+$ electron recombination rate. 
We also resolved a previous uncertainty in studies which considered the effects of dust grains  (cf. \citet{1995A&A...296..779S,1995A&A...297..251L,2006ApJ...645..314B,2006A&A...459..813W}) by showing  that, for realistic dense cloud model parameters (high sulfur depletion and reduced number of small grains), ion-grain recombination does not suppress the occurrence of  bistable solutions. 

The fact that two stable solutions can co-exist for the same set of physical and chemical parameters  may provide important insights into the observational interpretation  of interstellar  cloud chemistry (e.g. \citet{1997A&A...318..579G,2011ApJ...740L...4C}) and  perhaps other astronomical objects. 
All previous astrochemical models that have demonstrated bistable solutions have the oxygen autocatalysis as the cause. 
Hence, an important  question  is  whether or not other autocatalytic processes are present in interstellar chemistry?

Astrochemical reaction networks are large, complicated, and  highly nonlinear. Identifying any autocatalytic processes embedded within them can therefore be challenging. 
Based on our earlier work  (Paper I), one possible path to autocatalytic feedback and bistability is that the autocatalyst, X, be able for form a dimer, X$_2$. In interstellar chemistry,   direct autocatalysis through a dimer intermediate  (\citet{1975JChPh..62.1010T, 2011JPCA..115.8073P}) can occur in the sequence 
    \begin{eqnarray}
{\rm  X ~+~ XH  } ~& \longrightarrow&~ {\rm   X_2  ~+~ H}\label{x2}\\
{\rm   A^+   ~+~   X_2  } ~&\longrightarrow&~  {\rm  X_2^+  ~+~  A} \label{a+x2} \\
{\rm   X_2^+   ~+~   e^- } ~&\longrightarrow&~  {\rm   X ~+~ X} \label{x2++e}
\end{eqnarray}
      In pure oxygen chemistry X$_2$ is O$_2$, A$^+$ is H$^+$  or He$^+$,  and positive feedback   is provided through the sequence 
   \begin{equation} 
 {\rm O    + H_3^+  \longrightarrow  OH^+ \buildrel H_2 \over \longrightarrow H_2O^+ \buildrel H_2 \over
\longrightarrow H_3O^+  \buildrel e^- \over \longrightarrow OH } \label{oh-seq1}
\end{equation}
Autocatalysis can also occur   through
\begin{equation} 
{\rm  He^+   ~+~   O_2 ~  \longrightarrow  ~ O^+  ~+~  O  ~+~  He} \label{he++o2}.
\end{equation} 
followed by  the additional   feedback process 
 \begin{equation} 
 {\rm O^+    + H_2  \longrightarrow  OH^+ \buildrel H_2 \over \longrightarrow H_2O^+ \buildrel H_2 \over
\longrightarrow H_3O^+  \buildrel e^- \over \longrightarrow OH } \label{oh-seq1}
\end{equation}
Similar autocatalytic processes  occur in realistic cloud models, where S$^+$ produces two O atoms in a rapid 2-step reaction sequence involving SO$^+$, and C$^+$ reacts with O$_2$  to  produce O$^+$ and O in simultaneous product branchings  (Paper I). As nitrogen and  carbon chemistries also contain two simple dimers, N$_2$ and C$_2$,  we examined the pure systems for autocatalysis and bistability.

\section{Model calculation}  
\begin{table}[tbh]
        \begin{center}
        \caption{Dense Cloud Model Parameters }
        \begin{tabular}{lll}
        \hline
        \hline
        Temperature & $T$ & 10 K \\
        Visual extinction & $A_{\rm V}$ & 15 mag \\
        Cosmic ray ionization rate & $\zeta$ & $5.0\times 10^{-17}$ s$^{-1}$ \\
        \hline
        \end{tabular}
        \end{center}
\end{table}    
 To explore whether autocatalysis can occur via N$_2$ and C$_2$ we performed an analysis similar to that of Paper I.
 We used same physical parameters for all the models (Table 1), appropriate for  cold and dark cloud conditions. 
The chemical network comprises  136 species and 2001 reactions derived from the UMIST database (\citet{2013A&A...550A..36M}). 
The chemistry is driven by cosmic-ray (CR) ionization, gas-phase formation and destruction processes, with only molecular hydrogen  formed on the surfaces of dust grains. Neutralization of atomic ions on negatively charged grains occurs but, as shown in Paper I,  the surface area for ion-grain recombinations and the elemental S abundance  in dark clouds are both too low to have an effect on bistable solutions. Photoprocesses are neglected except for those resulting from  cosmic-ray-induced photons (\citet{1983ApJ...267..603P}).
  
As in many previous models, for simplicity we assume isothermality so that  $T $ is fixed at 10K, define $\zeta$, and vary $\zeta/n_ {\rm H}$.  We note that higher cosmic-ray ionization rates, e.g. $\zeta > 10^{-16}$ s$^{-1}$,  produce higher ionization fractions, but as they also heat the gas  a thermal balance calculation would then be required.

 
In Paper I,  H$_3^+$ electron recombination rate coefficient ($\rm \alpha_3$) was found to play a role in the cause of the oxygen chemistry bistability;  the calculations presented here also address this issue (section 3.2 and 3.3)
Although $ \alpha_3(T)$ is now very well constrained, both experimentally and theoretically (\citet{2003Natur.422..500M,2009PhRvL.102b3201J,2019RSPTA.37780397L}), 
  we have used two values that bracket the experimental value at 10K; $\rm \alpha_3(10K)$=$1.0  \times 10^{-6} \rm cm^{3}~s^{-1} $ \& $1.0  \times 10^{-7} \rm cm^{3}~s^{-1} $. Except where stated, all models were first computed using the higher $\rm \alpha_3$.
  These values were chosen to specifically  facilitate comparison with published bistability studies.  The model of Paper I was based on the bistability study of \citet{2003A&A...399..583C} which adopted  
an $\rm \alpha_3(10K)$ value of  $1.6  \times 10^{-6} \rm cm^{3}~s^{-1}$. This is ten times the rate coefficient of $3.00 \times 10^{-8}   T_3^{-0.5}$ $\rm cm^{3}~s^{-1}$  given by the UMIST database, then current as of \citet{1997A&AS..121..139M}, and chosen so that the solutions could be compared to those found in the bistability models published up to that time.  These earlier studies had $\rm \alpha_3(10K)$ values of  $8.2 \times 10^{-7} \rm cm^{3}~s^{-1}$  (\citet{1993ApJ...416L..87L}) and $1.0  \times 10^{-6} \rm cm^{3}~s^{-1}$ (\citet{1998A&A...334.1047L,1992MNRAS.258P..45P}). These 10K rate coefficients are  higher than that obtained from the fit to \citet{2003Natur.422..500M} rate coefficient given in the most recent UMIST database compilation,  $6.70 \times 10^{-8} (T/300 \rm K)^{-0.52} \rm cm^{3}~s^{-1}$ (\citet{2013A&A...550A..36M}). Other bistability studies have employed a lower 10K rate coefficient more in line with the McCall et al. value:  $1.6  \times 10^{-7} \rm cm^{3}~s^{-1}$  (\citet{2001A&A...370..557V}); $1.7  \times 10^{-7} \rm cm^{3}~s^{-1}$   (\citet{2006ApJ...645..314B}, at 50K); $3.0  \times 10^{-7} \rm cm^{3}~s^{-1}$  (\citet{2006A&A...459..813W}).


To demonstrate that bistability is an intrinsic property of the nitrogen and carbon chemistry,    we considered separately models employing  two  reduced networks. Model A, for the  nitrogen chemistry,  contains only the elements H, He and N;  Model B for the carbon chemistry  has only H, He and C. Steady-state chemical bifurcation diagrams were then computed as a function of  the hydrogen nucleon density,  $n_{\rm H}$,  across a range typical of values found in the  diffuse and  dense interstellar medium.


 The effect on the abundance profiles  of reducing the elemental abundances of  nitrogen  and  carbon from their undepleted interstellar values, ab$_N^0$ and ab$_C^0$, was explored.  For completeness, we also calculated the dependence of the oxygen bistability on elemental depletion. We assumed  values for ab$_O^0$ and ab$_C^0$ equal to  $8.53\times 10^{-4}$ ($\delta_1$) and  $3.62\times 10^{-4}$ ($\delta_1$), respectively (\citet{1989GeCoA..53..197A}).  For ab$_N^0$ we assumed $7.5\times 10^{-5}$ ($\delta_1$) (Meyer et al. 1997). For each of the chemistries we calculated the bifurcation behavior for depletion factors of unity $\delta_1$, 10 ($\delta_{0.1}$) and 100 ($\delta_{0.01}$); for the nitrogen chemistry (Model A) we  considered the additional case of a depletion factor of 2.0 ($\delta_{0.5}$). 


The steady-state chemical abundances in the models were found using  two different numerical methods : a time-dependent program using a DVODE solver for stiff differential equations and a steady-state program using Newton-Raphson iteration, to determine the unstable solutions.


\section{Bifurcation analysis}  
\begin{sidewaysfigure*}
\begin{center}
\includegraphics[width=0.70\textwidth]{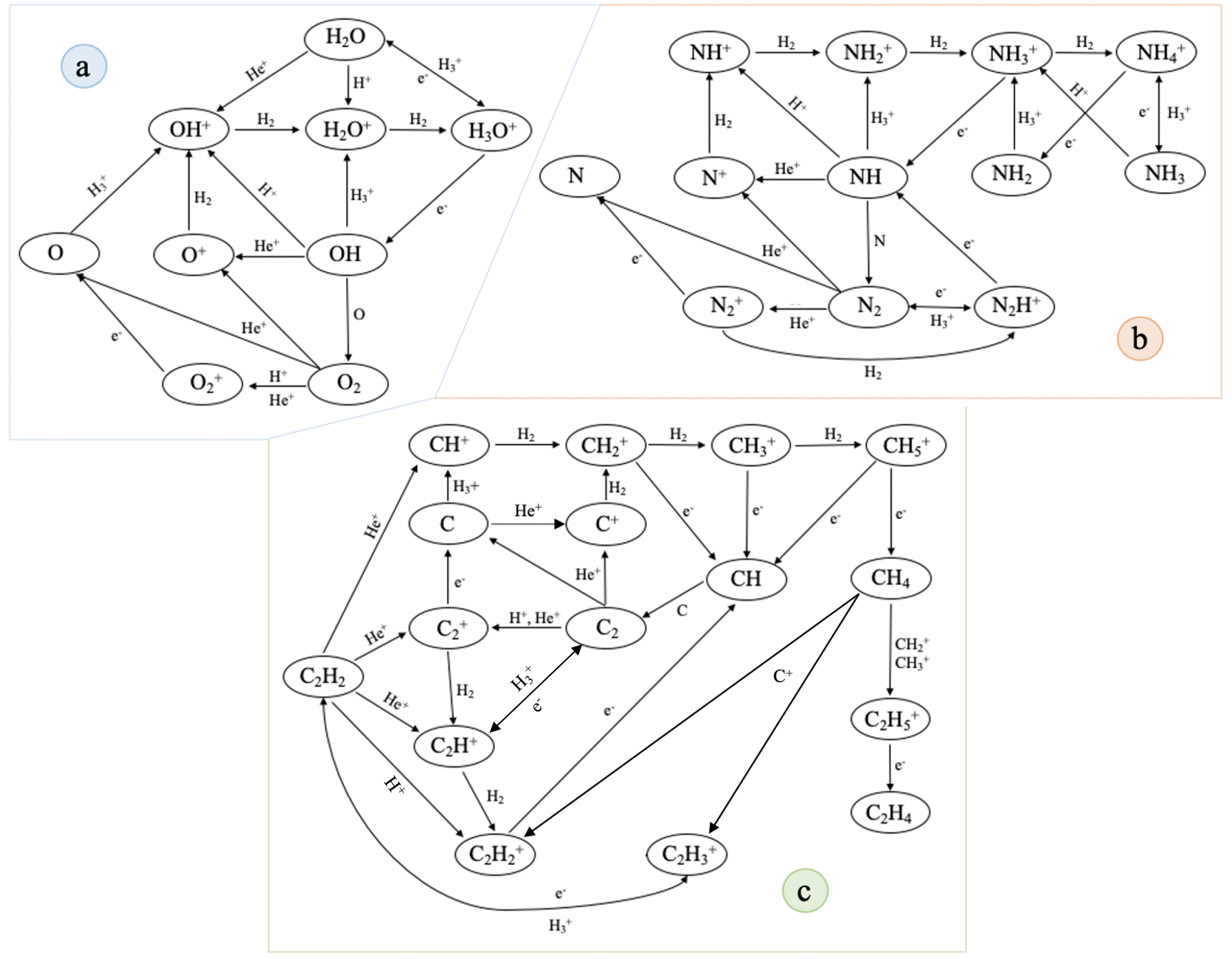}   
\caption{Schematic diagram for the separate chemistry of each of  (a) Oxygen, (b) Nitrogen and (c) Carbon. Only reactions potentially  important for autocatalysis are shown; several species (e.g. C$_2$H) and many ion-molecule and electron recombination processes have been omitted for clarity.}
\label{fig1_3}
\end{center}
\end{sidewaysfigure*}

In this section, we show how  bistable solutions occur when the  elemental depletions are considered as  control parameters, and identify the important  chemical reaction sequences responsible for the nonlinear behavior. { Figure \ref{fig1_3}} summarizes the key reactions in the basic gas phase oxygen, nitrogen and carbon chemistries of dense interstellar clouds. For the nitrogen and carbon networks many reactions have been omitted for clarity.

\subsection{Effect of elemental depletion}
 
\begin{figure*}[h]
      \includegraphics[width=\textwidth]{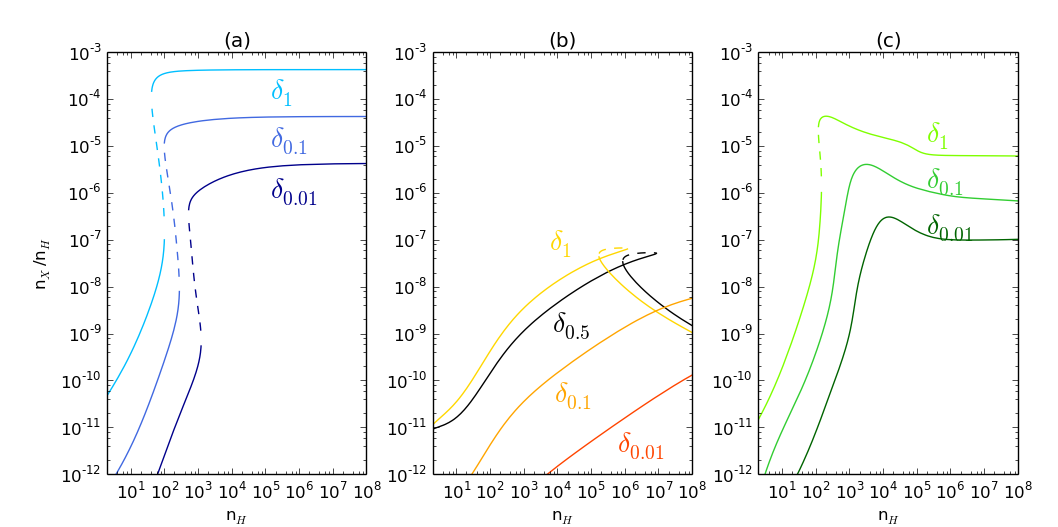}   
\caption{Fraction abundances as a function of density of (a) molecular oxygen, (b) molecular nitrogen and (c) molecular carbon for different depletion of ab$_O^0$, ab$_N^0$, ab$_C^0$. Note that full and broken curves respectively denote stable and unstable solutions.
\label{fig2_3}} 
 \end{figure*}

 Figure \ref{fig2_3} shows the calculated fractional abundances ($y_{\rm X}$ = $n({\rm X})$/$n_{\rm H}$) of N$_2$, C$_2$ and  O$_2$  as a function of $n_{\rm H}$  for successively larger  elemental depletions,  $\delta_m$, as described above.  In Figure \ref{fig2_3}(a) an  unstable branch appears in the nitrogen chemistry at all values of $\delta_m$. The morphology of the bifurcations appear as loops but are in fact the usual hysteresis curves projected  on to the  $\lbrack y_{\rm X} - n_{\rm H} \rbrack$ plane (see Figure 3 of \citet{2003A&A...399..583C}).
 These start to appear at densities  of  $n_{\rm H}$ $\sim$ a few$\times 10^5$ cm$^{-3}$, much higher than found for the corresponding oxygen model (Figure \ref{fig2_3}(c)), and persist up to very high densities.   The bifurcations in the $\delta_{0.1}$ and the $\delta_{0.01}$ models (not shown) occur at densities higher than  $10^8$ cm$^{-3}$ and $10^{10}$ cm$^{-3}$, respectively.  
   
  The reason that the N$_2$ abundance in these models is much smaller than found in complete chemical models is due to the omission of O and C from this reduced model, and the important N$_2$ formation pathways are absent, viz.

\begin{equation}
{\rm N ~+~ OH ~  \longrightarrow  ~ NO  ~+~ H}\label{n+oh}
\end{equation}
\begin{equation}
{\rm N ~+~ NO ~  \longrightarrow  ~  N_2  ~+~ O}\label{n+no}
\end{equation}

\begin{equation}
{\rm N ~+~ CN ~  \longrightarrow  ~  N_2  ~+~ C}\label{n+cn}
\end{equation}

Nevertheless, as discussed in  $\S 4$, the large atomic/molecular nitrogen ratios in these models could be relevant to conditions in prestellar cores.

   
Figure \ref{fig2_3}(b) shows that a   bifurcation in the C$_2$ abundance is only present in the reduced carbon chemistry when elemental carbon is undepleted. This occurs only at  $n_{\rm H}$ $\sim$ $10^2$ cm$^{-3}$ and only unique steady-state solutions are found at higher densities as  the carbon depletion is reduced to values  that are more relevant for dense clouds.
     In these reduced models,  C$_2$ dimers,   and other hydrocarbons (see Figure \ref{fig3_3}(d)), attain large abundances because of the absence of destructive oxygen atom reactions and the inability to form CO through OH and O$_2$.    In fact, we have discovered that C$_2$ is {\it not} the dimer associated with this bistability (see $\S3.3$).
    
    
The low density O$_2$ bifurcations for the reduced oxygen chemistry are shown in   { Figure \ref{fig2_3}(c)} and  are similar to those presented in \citet{2019ApJ...887...67D} (their Models 5 \& 6) for a single value of the elemental oxygen abundance ($8.53\times 10^{-5}$).  The important result here is that bifurcations exist for all values of $\delta_m$. This demonstrates that oxygen chemistry in dense molecular clouds is {\it inherently}  bistable and may explain why it was most likely to first appear in chemical models with significant depletion of the elements or low C/O ratios (\citet{1993ApJ...416L..87L,1998A&A...334.1047L}; Paper I). 
    
\subsection{The nitrogen chemistry}    
 
\begin{figure*}[h]
\begin{center}
   \includegraphics[width=\textwidth]{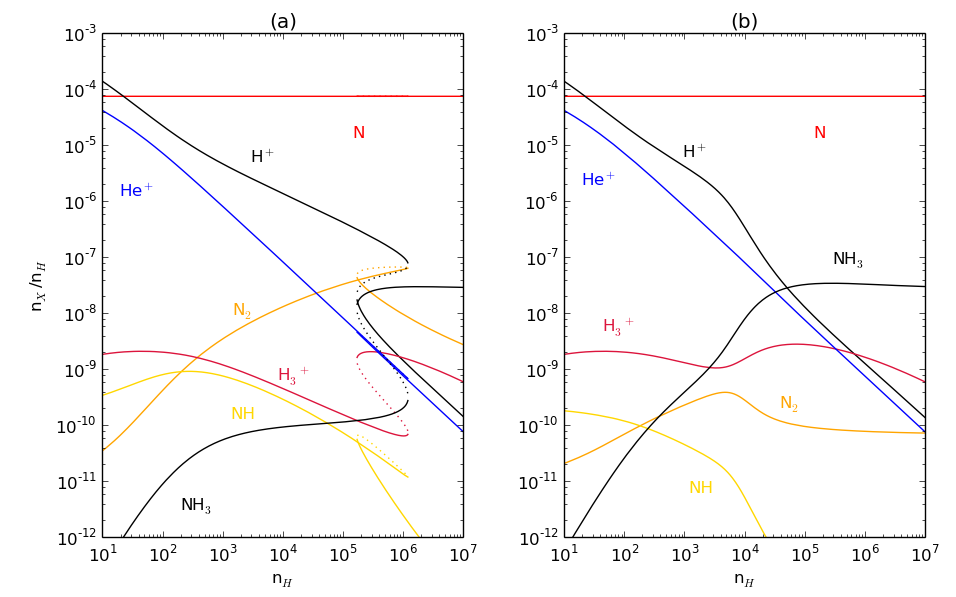}   
\caption{Fractional abundances as a function of density of the main species in the nitrogen chemistry of  Model A.    Figure (b) is as (a) except  that reaction (\ref{nh3+e-}) has been  artificially removed.} 
\label{fig3_3} 
\end{center}
 \end{figure*}

We identify the fundamental processes that control the occurrence of bistable solutions as autocatalysis in the pure nitrogen chemistry of Model A.
The relevant  nitrogen chemistry undergoes a series of reactions analogous to that found for oxygen in Paper I.
   \begin{eqnarray}
{\rm  N ~+~  NH } ~& \longrightarrow&~ {\rm N_2 ~+~  H.}\label{n+nh}\\
{\rm  He^+ ~+~ N_2 } ~&\longrightarrow&~  {\rm  N_2^+  ~+~ He }\label{he+n2} \\
{\rm   N_2^+   ~+~   e^- } ~&\longrightarrow&~  {\rm  N ~+~ N} \label{n2++e} 
\end{eqnarray}
except that H$^+$ does not form N$_2^+ $ by charge transfer to N$_2$. 
Paper I discussed the effect of varying the H$_3^+$ electron recombination rate, changing the efficiency of certain reactions responsible for bistability in the oxygen chemistry. However, unlike atomic oxygen, H$_3^+$ does not react with nitrogen atoms (e.g. \citet{2019A&A...625A..74R}; Figure \ref{fig2_3}(b)), i.e. 
  \begin{equation} 
 {\rm N ~+~  H_3^+   \longrightarrow  ~ no ~  reaction } \label{n+h3+} 
\end{equation}
 and so variation of  $\rm \alpha_3$ will have no effect on the appearance or disappearance of bistable solutions in the nitrogen chemistry.
    This was confirmed in calculations with  $\rm \alpha_3(10K)$=$1.0  \times 10^{-7} \rm cm^{3}~s^{-1} $; these produced identical results to those of Figure \ref{fig3_3} (not shown).
  Although positive feedback to form NH can occur through the sequence 
     \begin{eqnarray}
{\rm   N_2^+   ~+~   H_2  } ~& \longrightarrow&~ {\rm   N_2H^+ ~+~  H} \label{n2++h2} \\
{\rm N_2H^+ ~+~  e^-  } ~& \longrightarrow&~ {\rm  N_2  ~+~  H}\label{n2h+eA} \\
{\rm   } ~&\longrightarrow&~  {\rm  N  ~+~  NH}\label{n2h+eB}     
\end{eqnarray}
we have found that autocatalysis through reactions ({\ref{n+nh}}) - ({\ref{n2h+eB}}) does not produce the  bistable solutions found in the pure nitrogen models considered here.

Reaction with  He$^+$ initiates the most efficient positive feedback
\begin{equation}
{\rm He^+ ~+~ N_2 ~  \longrightarrow  ~ N^+ ~+~  N ~+~ He }\label{he++n2}
\end{equation}
The nitrogen ion then reacts with H$_2$ to produce NH$^+$ and the NH$_n^+$ ions after repetitive hydrogenation as shown in { Figure \ref{fig1_3}(b)}. The  hydride is formed in the electron dissociative recombination channel
\begin{equation}
{\rm NH_3^+ ~+~  e^- ~  \longrightarrow  ~ NH ~+~  H ~+~  H}\label{nh3+e-}
\end{equation}
so that   NH  is recovered for reaction  (\ref{n+nh}).  This autocatalytic process is favored  because the reaction 
\begin{equation}
{\rm NH_3^+ ~+~  H_2~  \longrightarrow  ~ NH_4^+ ~+~  H}\label{nh3+h2}
\end{equation}
has a  slow reaction rate coefficient compared to those of NH$^+$ and NH$_2^+$ with H$_2$ (e.g. \citet{2013A&A...550A..36M,2019A&A...625A..74R}) and this allows NH$_3^+$  electron recombination in reaction (\ref{nh3+e-}) to be competitive with proton transfer.  We determined that no other NH$_n^+$-electron recombinations were crucial for bistability  in Model A.
 
Figure \ref{fig3_3} presents the computed chemical abundances for Model A.  As found in previous studies, the bifurcation  divides the solutions between a high ionization phase dominated by H$^+$ and He$^+$,  the HIP, and a  low ionization phase where H$_3^+$ is the more abundant ion, the LIP (\citet{1993ApJ...416L..87L}). Figure \ref{fig3_3}(a) shows that,  for the case of an N elemental abundance of $7.5 \times 10^{-5}$ (i.e. $\delta_{1}$), the abundance profile of N is  highest followed by N$_2$ and  NH$_3$.  As noted above,  the N$_2$ formation is inhibited because OH and CN are absent in this model. 
We can confirm the autocatalytic nature by artificially removing key reactions. For example,   if  the rate coefficient of reaction (\ref{nh3+e-}) is set to zero, bistability in the system disappears, as shown in Figure \ref{fig3_3}(b). In Model A, we find that autocatalysis  initiated by  He$^+$ (via reaction (\ref{he++n2})) is  solely  responsible for bistability, and that other processes that could break down N$_2$ - UV photolysis or N$_2^+$ recombination - are not competitive.

\subsection{The carbon chemistry} 

 In  carbon chemistry  direct autocatalysis through the C$_2$  dimer intermediate is possible, similar to that of the oxygen chemistry (see Figure \ref{fig1_3}).   Neutral-neutral reaction of atomic carbon with the methylidyne radical  (CH) forms C$_2$ 
\begin{equation}
{\rm C ~+~ CH ~  \longrightarrow  ~ C_2  ~+~ H.}\label{c+ch}
\end{equation} 
 and reactions with H$^+$ and He$^+$  provide autocatalysis 
\begin{equation} 
{\rm  H^+   ~+~   C_2 ~  \longrightarrow  ~ C_2^+  ~+~  H} \label{h++c2} 
\end{equation}  
\begin{equation} 
{\rm  He^+   ~+~   C_2 ~  \longrightarrow  ~ C_2^+  ~+~  H ~+~  He} \label{he++c2} 
\end{equation}  
\begin{equation} 
{\rm    C_2^+   ~+~   e^- ~  \longrightarrow  ~ C ~+~ C} \label{c2++e} 
\end{equation} 
 The 2C produced initiate an ion-molecule feedback  sequence culminating in CH formation by electron dissociative recombination of CH$_2^+$, CH$_3^+$ and CH$_5^+$
\begin{equation} 
 {\rm C ~+~  H_3^+   \longrightarrow  CH^+ \buildrel H_2 \over \longrightarrow   ~ ... \longrightarrow  ~  CH_n^+  \buildrel e^- \over \longrightarrow CH } \label{ch-seq1} 
 \end{equation} 
 CH is also produced in 
  \begin{equation} 
 {\rm C^+ ~+~  H_2   \longrightarrow  CH_2^+ \buildrel H_2\over \longrightarrow  ~ ...  \longrightarrow  ~  CH_n^+  \buildrel e^- \over \longrightarrow CH } \label{ch-seq2}  
\end{equation}
where the  C$^+$ arises from 
\begin{equation}
{\rm He^+ ~+~ C_2 ~  \longrightarrow  ~ C^+ ~+~  C ~+~  He}\label{he++c2}
\end{equation}
 
 

\begin{figure*}
\begin{center}
   \includegraphics[width=\textwidth]{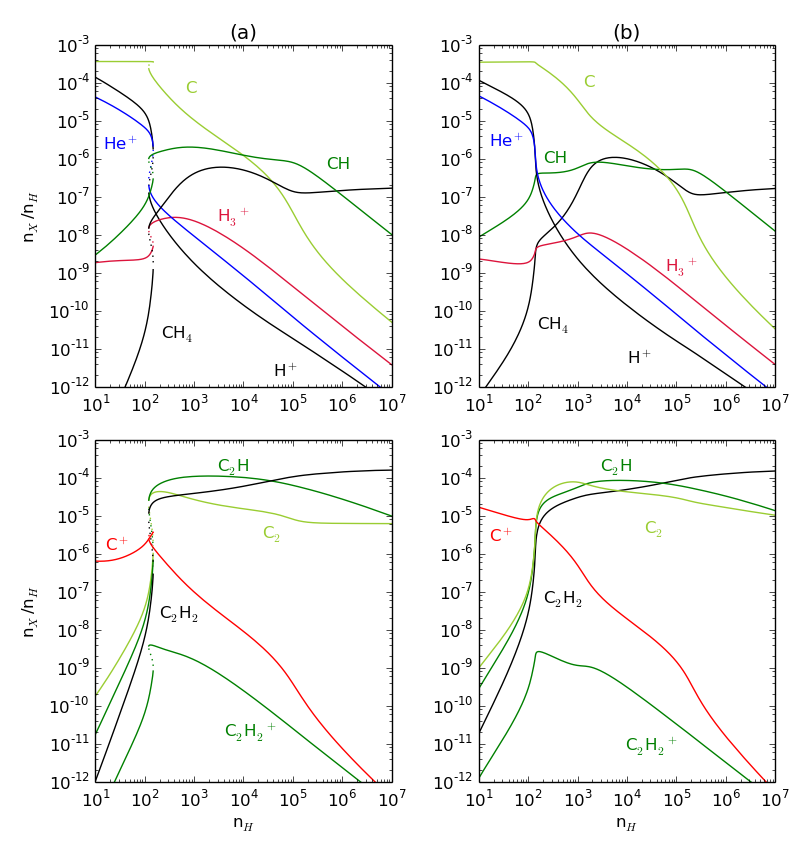}   
\caption{Fractional abundances as a function of density of the main species in the carbon chemistry of Model B.   Figure (b) is as (a) except  that reaction (\ref{c2h2++e-}) has been   artificially removed}
\label{fig4_3} 
\end{center}
\end{figure*}

Figure \ref{fig4_3} shows that,  for the case of a C elemental abundance of $3.62 \times 10^{-4}$ (i.e. $\delta_{1}$), atomic and molecular carbon (C$_2$), ethynyl (C$_2$H) and acetylene (C$_2$H$_2$) are the main chemical elements present along the HIP and the LIP  in  the carbon chemistry of Model B.
However, although the autocatalytic processes summarized above do occur, calculations in which reactions are selectively removed (not shown) have demonstrated they are not responsible for the bistable solution and the underlying  bifurcation shown in Figures \ref{fig2_3}(b) and \ref{fig4_3}(a).
 Instead, we have found that other autocatalytic processes are present in carbon chemistry and that these become dominant in the chemical network.
 In contrast to O$_2^+$,  C$_2^+$ can react with H$_2$, effectively bypassing reaction ({\ref{c2++e}) to form  C$_2$-hydrocarbon ions 
 (Figure \ref{fig1_3}(c))}\footnote{    Note that although the reaction $ {\rm  C_2H_2^+   ~+~   H_2 ~  \longrightarrow  ~ C_2H_3^+ ~+~   H }$  is listed in the current UMIST database with a 10K rate coefficient of  $1.0  \times 10^{-11} \rm cm^{3}~s^{-1}$  (\citet{2013A&A...550A..36M}), it has been removed from our chemical network as it is endothermic (\citet{2005PCCP....7.1592S}).}

 \begin{equation} 
{\rm    C_2^+   ~+~   H_2 ~  \longrightarrow  ~ C_2H^+ ~+~  H} \label{c2++h2} 
\end{equation}  
\begin{equation} 
{\rm    C_2H^+   ~+~   H_2 ~  \longrightarrow  ~ C_2H_2^+ ~+~   H} \label{c2h++h2} 
\end{equation} 
 \begin{equation}
{\rm C_2H_2^+ ~+~ e^- ~  \longrightarrow  ~ CH ~+~  CH}\label{c2h2++e-}
\end{equation}

Electron recombination of the acetylene ion ($\rm C_2H_2^+$) in reaction (\ref{c2h2++e-})  provides autocatalysis;  this is confirmed in {Figure \ref{fig4_3}(b)} which shows that the bistable solution disappears when the rate coefficient of this reaction is set to zero. Thus, CH  takes the part of the autocatalyst and  $\rm C_2H_2$  (CH$\equiv$CH)  becomes the related intermediate neutral dimer.
 Acetylene is formed in the reactions
 \begin{eqnarray}
\rm C^+   ~+~   CH_4 ~  & \longrightarrow & \rm ~ C_2H_3^+ ~+~   H \label{c++ch4a}\\
				& \longrightarrow  &   \rm  ~ C_2H_2^+ ~+~  H_2 \label{c++ch4b} 
\end{eqnarray}
\begin{equation} 
{\rm    C_2H_3^+   ~+~   e^- ~  \longrightarrow  ~ C_2H_2 ~+~  H} \label{c2h3++e} 
\end{equation} 

but is not crucial for this particular autocatalysis as, although $\rm C_2H_2$ can form $\rm C_2H_2^+$ by charge transfer from H$^+$ and He$^+$, the latter ion can form  independently of its presence, as above.


\begin{sidewaysfigure*}
\begin{center}
\includegraphics[width=0.75\textwidth]{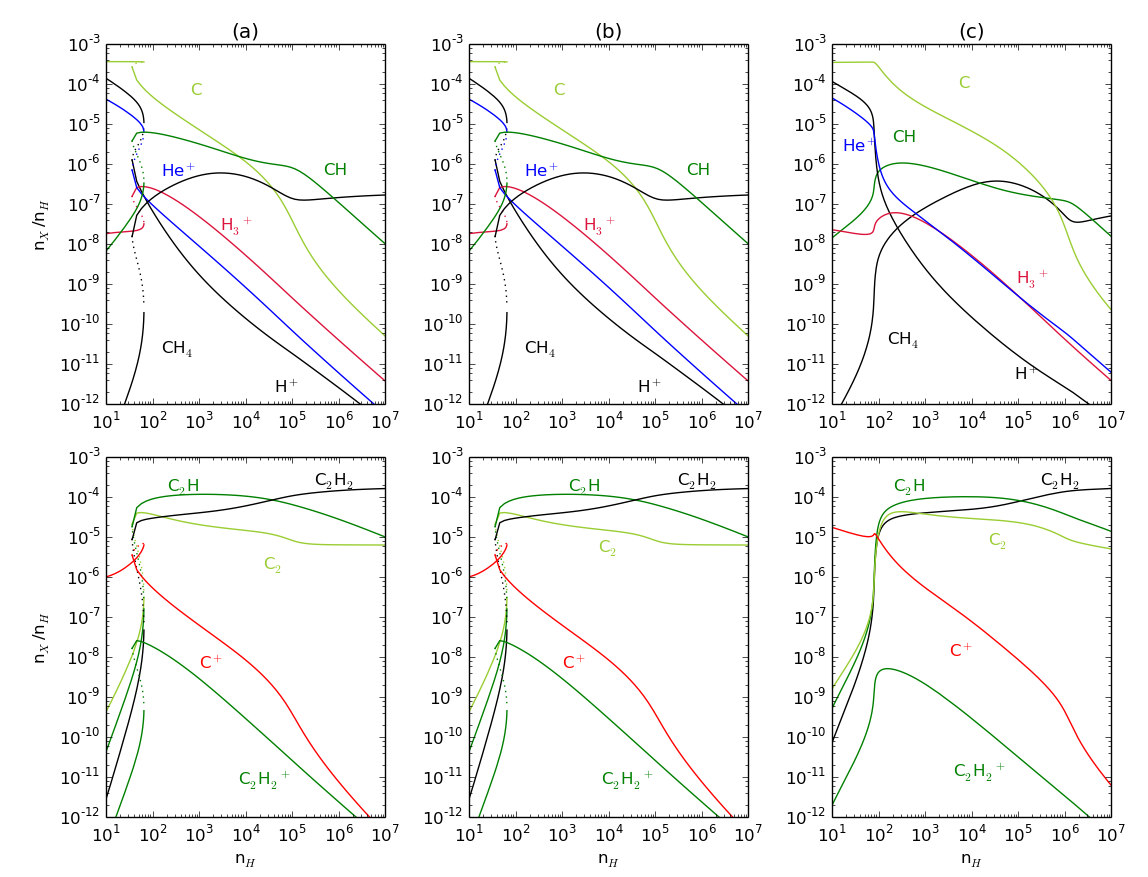}   
\caption{The carbon chemistry of Model B as in Figure 4 but with 
  $\alpha_3 = 1\times10^{-7}$ cm$^3$s$^{-1}$.   Figure (b) is as (a) except  that reaction (\ref{c2h2++e-}) has been removed.   Figure (c) is as (a) except  that  reactions (\ref{he++c2h2b}), (\ref{he++c2h2c}) and (\ref{he++c2h2d}) have been removed.}
\label{fig5_3} 
\end{center}
 \end{sidewaysfigure*}

As the H$_3^+$ electron recombination rate, $\rm \alpha_3$,  plays a role in setting  the electron number density in the LIP, and the  formation rates of CH$_n^+$ ions and CH$_4$ through the first step in sequence  (\ref{ch-seq1}),   we explored its effect on the carbon chemistry bifurcation. 
Figure \ref{fig5_3}(a)  shows that bistability is evident   in  the carbon chemistry of Model B with $\rm \alpha_3 = 1\times10^{-7}$ cm$^3$s$^{-1}$).   
This bifurcation occurs at lower $n_{\rm H}$  and Figure \ref{fig5_3}(b) shows that it  is still present even when the key autocatalytic reaction (\ref{c2h2++e-}) identified above is removed,  indicating that at the lower $\rm \alpha_3$  another, different,  autocatalytic mechanism is operating.
We have identified the relevant autocatalytic process as beginning with destruction of the acetylene dimer by  He$^+$   
  \begin{eqnarray}
\rm He^+ ~+~ C_2H_2  ~ & \longrightarrow & \rm ~ C_2H_2^+ ~+~ He \label{he++c2h2a}\\
				& \longrightarrow  &   \rm  ~ C_2^+ ~+~  H_2 ~+~ He  \label{he++c2h2b}\\
				& \longrightarrow  &   \rm  ~ C_2H^+ ~+~  H ~+~ He \label{he++c2h2c}\\
				& \longrightarrow  &   \rm  ~ CH^+ ~+~  CH ~+~ He  \label{he++c2h2d}
\end{eqnarray}
Reaction (\ref{he++c2h2a}) contributes to the autocatalysis of { Figure \ref{fig4_3}(a)} but the three dissociative charge transfer channels are crucial for the bistable solution of { Figures \ref{fig5_3}(a) \& (b)}.  This is confirmed in { Figure \ref{fig5_3}(c)}  which shows that,   in an artificial  version of Model B with the lower $\rm \alpha_3$ value,  no bifurcation  occurs once the rates of all three of reactions   (\ref{he++c2h2b}) - (\ref{he++c2h2d}) are set to zero. 
Reaction (\ref{he++c2h2d})  produces one CH molecule directly, and injection of the $\rm CH^+$ ion into the rapid  ion-molecule sequence leading to the CH$_n^+$ ions eventually  leads to a second (see Figure \ref{fig1_3}(c)).  The  $\rm C_2^+$ and  $\rm C_2H^+$ contribute to $\rm C_2H_2^+$ formation and  the eventual production of CH in reaction  (\ref{c2h2++e-}),  whereas   reaction (\ref{c++ch4a})  contributes  feedback to $\rm C_2H_2$  reformation. 
Thus, acetylenic autocatalysis contains processes analogous to  the oxygen autocatalysis : destruction of the dimer ion ($\rm O_2^+$ or $\rm C_2H_2^+$) and $\rm He^+$ attack on the neutral  ($\rm O_2$ or $\rm C_2H_2$).   
 The bistable solutions only occur at low densities and high ionization  because the key processes rely on competition between electron recombinations and reactions with H$_2$. This is also the reason why bistability only occurs in the model with the highest carbon abundance; relative to  O$^+$ and N$^+$,  C$^+$ reacts very slowly with H$_2$ by radiative association and so can be an additional source of electrons (e.g. Figure \ref{fig4_3}).


  
\section{Discussion } 

To more easily understand their autocatalytic potential, we followed the approach  of Paper I and deliberately decoupled the C and N chemistries into their basic reaction networks.
In Paper I we showed how the basic oxygen chemistry of { Figure \ref{fig1_3}(a)} produced bistability  in a realistic molecular cloud chemistry as addition of elemental C and S,  and variation of other control parameters,  moved the bistable region around in the $\lbrack y_{\rm X} - n_{\rm H} \rbrack$ plane. This was straightforward because we started from a known bistable solution.  
Re-coupling the reduced oxygen, carbon and nitrogen chemistries  and understanding  how the O, C and N autocatalytic cycles influence, and are influenced by, each other in molecular clouds  is  not trivial.  
For example, the fact that  OH catalyses N$_2$ formation in molecular clouds 
\begin{equation}
  \rm{ N 
  \buildrel OH \over \longrightarrow NO
  \buildrel N  \over \longrightarrow N_2 }
  \label{ncycle}
\end{equation}
also means that N atoms could influence bistable solutions occurring oxygen chemistry.  
   Furthermore,  we know from Paper I that re-coupling  carbon and sulfur chemistry to the  oxygen  chemistry 
 will tend to move their bistable regions   to  higher $n_{\rm H}$ at fixed $\zeta$. \footnote{
    Although S$^+$ can copiously provide electrons, and it is  known that the reaction of S$^+$ with O$_2$ can lead to bistable solutions (\citet{2006ApJ...645..314B}; Paper I),  recent estimates of the gaseous S abundance in TMC-1 (\citet{2019A&A...624A.105F}) suggests relatively large depletion in dense clouds.  However, since S$_2$ is a viable dimer, we did investigate whether or not a reduced He-H-S network could produce bistability. No such solutions were found. This is most likely because of the inefficiency of reaction sequences leading to (re)formation of S$_2$ due to the endothermicity of S$^+$  and SH$^+$  reactions with H$_2$ (e.g. \citet{2013A&A...550A..36M}). 
 } 
Thus it is possible that  parameter variations exist whereby multiple bistable regions can co-exist in the $\lbrack y_{\rm X} - n_{\rm H} \rbrack$ plane, leading to exchange of autocatalysts between coupled bistable systems that in turn can lead to oscillations and chaos ((e.g.  \citet{1996OxPres...314...G,2003DaltTra...7..1201S}).  
Such a study is  beyond the scope of this paper and will be presented elsewhere.
        
%
   

The bifurcation analysis presented here and in Paper I  is based on steady-state chemical abundances.
However, it is known that  bifurcations in the abundances occurs much earlier, on time-scales relevant for molecular cloud life-times (\citet{1995A&A...297..251L}), and so, although found in reduced reaction networks, it is of interest to see if these new bistable solutions  could be applicable to actual astronomical environments.    

   In cold, dense prestellar  cores ($n_{\rm H}$ $\sim$   $10^5-10^8 \rm cm^{-3} $), where  CO is   observed to be selectively depleted on dust  relative to N$_2$ (\citet{2007ARA&A..45..339B}), the nitrogen chemistry that occurs is identical to that of Model A 
 (\citet{2002ApJ...569L.133C}) and so  bistable solutions are possible at the densities of interest.   This environment could  be conducive for  producing the enhanced $\rm ^{14}N/^{15}N$ isotopic ratios found in dark clouds,  comets and meteorites (e.g. \citet{2016IAUFM..29A.271W,2018MNRAS.474.3720W,2018ApJ...857..105F}).
Based on Model A, any $^{15}$N enrichment would occur in ammonia and related hydrides  (\citet{2002ApJ...569L.133C}). Apart from temperature, the rate of the reaction that initiates the ion-molecule feedback  
    \begin{equation}
{\rm N^+ ~+~  H_2 ~  \longrightarrow  ~ NH^+ ~+~  H.}\label{n++h2}
\end{equation}
       depends on the spin-state of  H$_2$ through the ortho/para  ratio (OPR, e.g. \citet{1991A&A...242..235L,2012A&A...537A..20D}). This  introduces the H$_2$  OPR as another control parameter for bistable solutions, one  which also directly affects the $^{15}$N  fractionation kinetics (\citet{2012ApJ...757L..11W}).
    Thus, the fact that  two possible solutions can exist suggests that a re-examination of  chemical models for dense cloud nitrogen fractionation is warranted.
          
Following submission  of the original manuscript we became aware of the study of  \citet{2020A&A...643A.121R}  reporting the discovery of limit cycle oscillations in gas-phase interstellar chemistry models. 
Roueff \& Le Bourlot  found that the oscillations were present in the nitrogen chemistry and that their onset was in fact controlled by the H$_2$ OPR.  During revision of the manuscript we explored whether Hopf bifurcations could develop in our reduced N chemistry bifurcation diagrams by introducing the H$_2$ OPR as an additional control parameter. We repeated the reduced N model calculations underlying Figure \ref{fig3_3} but found that the crucial range of OPR values found by Roueff \& Le Bourlot did not produce Hopf bifurcations.

 
The bistable solutions  found in  the carbon chemistry require high electron fractions and low values of both the H$_2$ density and the elemental carbon depletion and so are unlikely to be applicable to dense molecular clouds. One environment that could possibly be conducive to  carbon chemistry autocatalyses reported here is Titan's N$_2$-CH$_4$ ionosphere, where many large organic molecules are known to be present (\citet{2017JGRE..122..432H}). In fact, bistable solutions have already been reported in a sensitivity analysis of Titan photochemical models (\citet{2008P&SS...56.1630D}). It was found that the chemistry of C$_2$H$_2$ and C$_2$H$_4$ could be bistable although the underlying chemical pathways, involving neutral-neutral processes, differ from those studied here. The physical environment of Titan is significantly different from that of dense interstellar clouds,  nevertheless, as H$_2$ is much less abundant on Titan,  this work suggests that Titan's ion-molecule chemistry could also lead to bistable solutions.

 Finally,  the interstellar  chemistry of carbon is much more complex than studied here, where we have only considered up to C$_2$-hydrocarbons.
Autocatalytic processes in addition to those described above are likely present. For example, ethene, when viewed as a dimer (CH$_2$)$_2$ takes part in autocatalysis for CH$_2$ in reaction sequences similar to those for acetylene.  Although present in our network, we found this played no role in producing bistable solutions. 
Analysis of larger, more extensive, carbon chemistry networks may lead to the discovery of more deeply embedded, and less so apparent, autocatalytic cycles.   If one identifies the higher hydrocarbons as dimers then diacetylene  (C$_4$H$_2$) and the carbon tetramer (C$_4$) could  act as respective  intermediates  for C$_2$H and C$_2$ autocatalysis.  
The possibility for autocatalysis involving larger organic molecules remains to be explored;  anion chemistry (e.g. \citet{2012ApJ...749..120C,2013A&A...550A..36M}) may become as  important as the cationic reactions considered here. 
      
 \section{Conclusions} 
  
We have found new bistable solutions in reduced chemical sets comparable to simplistic gas phase chemical models of dense molecular clouds.
We have identified several  autocatalytic cycles in the ion-molecule chemistry of the major volatile elements: O, N and C.
In nitrogen chemistry there are two possible autocatalytic processes. An  autocatalytic cycle driven by the reaction of He$^+$ with N$_2$ produces bistability. We  find that this model has bistable solutions at high densities  ($n_{\rm H} > 10^5 \rm cm^{-3} $) which are present at all values of the elemental N abundance considered. 

The bistable solutions  in the reduced model of the oxygen chemistry occur at low densities but are similarly present at all values  of elemental O depletion, including none. Thus, bistable solutions are a fundamental property of interstellar chemistry; introduction of other elements,  carbon in particular, acts to suppress these solutions by inhibiting autocatalysis through reactions with O$_2$  and OH.   

 The carbon chemistry exhibits  bistable solutions which appear around 10$^2$ cm$^{-3}$, or less, but only when an elemental depletion  equivalent to that of the diffuse ISM is adopted. We have identified three possible autocatalytic processes.  
  We find that two autocatalyses occur in the acetylene chemistry with  CH, rather than atomic C, being the autocatalyst. These involve $\rm C_2H_2^+$ and $\rm C_2H_2$ in processes analogous to those found in the oxygen autocatalysis; their relative influence is similarly determined by the $\rm H_3^+$ recombination rate.    \footnote{ We note that, as $ \alpha_3(T)$ is known experimentally to within 20\% (\citet{2003Natur.422..500M}), only bistable solutions with the appropriate 
  $\rm \alpha_3(10K) \approx 3.9   \times 10^{-7} \rm cm^{3}~s^{-1}$ will occur in realistic molecular cloud models.   }   As there are no  bistable solutions for the appropriate  densities ($n_{\rm H} > 2 \times 10^3\rm cm^{-3}$) and elemental carbon depletions, these bistable solutions are unlikely to be relevant for realistic  models of dense clouds.


The fact that several, and probably many, autocatalyses are embedded in gas-phase interstellar  chemistry, and yet only become active 
for certain combinations of the external control parameters,  is consistent with the definition  of  autocatalysis in general chemical systems given by  \citet{2011JPCA..115.8073P},  in which a distinction is made between {\it potential} and {\it effective} autocatalysis. The core of interstellar chemistry contains many key reactions that have been known for decades but whose autocatalytic potential was never identified  because the  autocatalytic processes where overwhelmed by the many other reactions in the chemical network. Only in  models reduced to the pure chemistry of  each of O, N and C do these autocatalyses become effective and apparent.

This work accounts for the simplest bistable solutions that can occur for  O, C and N chemistries in cold, dark molecular gas irradiated by cosmic rays; it is possible that more  bistable solutions can be discovered  through the coupling of these autocatalytic processes. 
The recent discovery of limit cycles by \citet{2020A&A...643A.121R} indicates that oscillations  and chaos may naturally arise in interstellar chemistry models that can exhibit multistability.


%

\vspace{1.0cm} 

We thank two anonymous referees whose comments significantly improved the manuscript.

This research was supported by the NASA Planetary Science Division Internal Scientist Funding Program through the Fundamental Laboratory Research work package (FLaRe) and by the NASA Goddard Science Innovation Fund. 

\bibliographystyle{aa}
\bibliography{biblio}

\begin{thebibliography}{42}
\expandafter\ifx\csname natexlab\endcsname\relax\def\natexlab#1{#1}\fi

\bibitem[{{Ag{\'u}ndez} \& {Wakelam}(2013)}]{2013ChRv..113.8710A}
{Ag{\'u}ndez}, M. \& {Wakelam}, V. 2013, Chem. Rev., 113, 8710

\bibitem[{{Anders} \& {Grevesse}(1989)}]{1989GeCoA..53..197A}
{Anders}, E. \& {Grevesse}, N. 1989, Geo. Cosmo. Acta, 53, 197

\bibitem[{{Bergin} \& {Tafalla}(2007)}]{2007ARA&A..45..339B}
{Bergin}, E.~A. \& {Tafalla}, M. 2007, \araa, 45, 339

\bibitem[{{Boger} \& {Sternberg}(2006)}]{2006ApJ...645..314B}
{Boger}, G.~I. \& {Sternberg}, A. 2006, \apj, 645, 314

\bibitem[{{Ceccarelli} {et~al.}(2011){Ceccarelli}, {Hily-Blant}, {Montmerle},
  {Dubus}, {Gallant}, \& {Fiasson}}]{2011ApJ...740L...4C}
{Ceccarelli}, C., {Hily-Blant}, P., {Montmerle}, T., {et~al.} 2011, \apjl, 740,
  L4

\bibitem[{{Charnley} \& {Markwick}(2003)}]{2003A&A...399..583C}
{Charnley}, S.~B. \& {Markwick}, A.~J. 2003, \aap, 399, 583

\bibitem[{{Charnley} \& {Rodgers}(2002)}]{2002ApJ...569L.133C}
{Charnley}, S.~B. \& {Rodgers}, S.~D. 2002, \apjl, 569, L133

\bibitem[{{Cordiner} \& {Charnley}(2012)}]{2012ApJ...749..120C}
{Cordiner}, M.~A. \& {Charnley}, S.~B. 2012, \apj, 749, 120

\bibitem[{{Dislaire} {et~al.}(2012){Dislaire}, {Hily-Blant}, {Faure}, {Maret},
  {Bacmann}, \& {Pineau Des For{\^e}ts}}]{2012A&A...537A..20D}
{Dislaire}, V., {Hily-Blant}, P., {Faure}, A., {et~al.} 2012, \aap, 537, A20

\bibitem[{{Dobrijevic} {et~al.}(2008){Dobrijevic}, {Carrasco}, {H{\'e}brard},
  \& {Pernot}}]{2008P&SS...56.1630D}
{Dobrijevic}, M., {Carrasco}, N., {H{\'e}brard}, E., \& {Pernot}, P. 2008,
  \planss, 56, 1630

\bibitem[{{Dufour} \& {Charnley}(2019)}]{2019ApJ...887...67D}
{Dufour}, G. \& {Charnley}, S.~B. 2019, \apj, 887, 67

\bibitem[{{Epstein} \& {Showalter}(1996)}]{1996ApJ...100..31E}
{Epstein}, I.~R. \& {Showalter}, K. 1996, J. Phys. Chem., 100, 13132

\bibitem[{{Fuente} {et~al.}(2019){Fuente}, {Navarro}, {Caselli}, {Gerin},
  {Kramer}, {Roueff}, {Alonso-Albi}, {Bachiller}, {Cazaux}, {Commercon},
  {Friesen}, {Garc{\'\i}a-Burillo}, {Giuliano}, {Goicoechea}, {Gratier},
  {Hacar}, {Jim{\'e}nez-Serra}, {Kirk}, {Lattanzi}, {Loison}, {Malinen},
  {Marcelino}, {Mart{\'\i}n-Dom{\'e}nech}, {Mu{\~n}oz-Caro}, {Pineda},
  {Tafalla}, {Tercero}, {Ward-Thompson}, {Trevi{\~n}o-Morales},
  {Rivi{\'e}re-Marichalar}, {Roncero}, {Vidal}, \&
  {Ballester}}]{2019A&A...624A.105F}
{Fuente}, A., {Navarro}, D.~G., {Caselli}, P., {et~al.} 2019, \aap, 624, A105

\bibitem[{{Furuya} \& {Aikawa}(2018)}]{2018ApJ...857..105F}
{Furuya}, K. \& {Aikawa}, Y. 2018, \apj, 857, 105

\bibitem[{{Gerin} {et~al.}(1997){Gerin}, {Falgarone}, {Joulain}, {Kopp}, {Le
  Bourlot}, {Pineau des Forets}, {Roueff}, \& {Schilke}}]{1997A&A...318..579G}
{Gerin}, M., {Falgarone}, E., {Joulain}, K., {et~al.} 1997, \aap, 318, 579

\bibitem[{{Gray} \& {Scott}(1996)}]{1996OxPres...314...G}
{Gray}, P. \& {Scott}, S.~K. 1996, J. Fluid Mech., 314, 406

\bibitem[{{H{\"o}rst}(2017)}]{2017JGRE..122..432H}
{H{\"o}rst}, S.~M. 2017, J. Geophys. R. (Planets), 122, 432

\bibitem[{{Jungen} \& {Pratt}(2009)}]{2009PhRvL.102b3201J}
{Jungen}, C. \& {Pratt}, S.~T. 2009, Phys. Rev. Lett., 102, 023201

\bibitem[{{Larsson}(2019)}]{2019RSPTA.37780397L}
{Larsson}, M. 2019, Philosophical Transactions of the Royal Society of London
  Series A, 377, 20180397

\bibitem[{{Le Bourlot}(1991)}]{1991A&A...242..235L}
{Le Bourlot}, J. 1991, \aap, 242, 235

\bibitem[{{Le Bourlot} {et~al.}(1995{\natexlab{a}}){Le Bourlot}, {Pineau des
  Forets}, \& {Roueff}}]{1995A&A...297..251L}
{Le Bourlot}, J., {Pineau des Forets}, G., \& {Roueff}, E. 1995{\natexlab{a}},
  \aap, 297, 251

\bibitem[{{Le Bourlot} {et~al.}(1995{\natexlab{b}}){Le Bourlot}, {Pineau des
  Forets}, {Roueff}, \& {Flower}}]{1995A&A...302..870L}
{Le Bourlot}, J., {Pineau des Forets}, G., {Roueff}, E., \& {Flower}, D.~R.
  1995{\natexlab{b}}, \aap, 302, 870

\bibitem[{{Le Bourlot} {et~al.}(1993){Le Bourlot}, {Pineau des Forets},
  {Roueff}, \& {Schilke}}]{1993ApJ...416L..87L}
{Le Bourlot}, J., {Pineau des Forets}, G., {Roueff}, E., \& {Schilke}, P. 1993,
  \apjl, 416, L87

\bibitem[{{Lee} {et~al.}(1998){Lee}, {Roueff}, {Pineau des Forets},
  {Shalabiea}, {Terzieva}, \& {Herbst}}]{1998A&A...334.1047L}
{Lee}, H.~H., {Roueff}, E., {Pineau des Forets}, G., {et~al.} 1998, \aap, 334,
  1047

\bibitem[{{McCall} {et~al.}(2003){McCall}, {Huneycutt}, {Saykally}, {Geballe},
  {Djuric}, {Dunn}, {Semaniak}, {Novotny}, {Al-Khalili}, {Ehlerding},
  {Hellberg}, {Kalhori}, {Neau}, {Thomas}, {{\"O}sterdahl}, \&
  {Larsson}}]{2003Natur.422..500M}
{McCall}, B.~J., {Huneycutt}, A.~J., {Saykally}, R.~J., {et~al.} 2003, \nat,
  422, 500

\bibitem[{{McElroy} {et~al.}(2013){McElroy}, {Walsh}, {Markwick}, {Cordiner},
  {Smith}, \& {Millar}}]{2013A&A...550A..36M}
{McElroy}, D., {Walsh}, C., {Markwick}, A.~J., {et~al.} 2013, \aap, 550, A36

\bibitem[{{Millar} {et~al.}(1997){Millar}, {Farquhar}, \&
  {Willacy}}]{1997A&AS..121..139M}
{Millar}, T.~J., {Farquhar}, P.~R.~A., \& {Willacy}, K. 1997, \aaps, 121, 139

\bibitem[{{Pineau Des For{\^e}ts} \& {Roueff}(2000)}]{2000RSPTA.358.2549P}
{Pineau Des For{\^e}ts}, G. \& {Roueff}, E. 2000, in Astronomy, physics and
  chemistry of H$^{+}$$_{3}$, Vol. 358, 2359--2559

\bibitem[{{Pineau des Forets} {et~al.}(1992){Pineau des Forets}, {Roueff}, \&
  {Flower}}]{1992MNRAS.258P..45P}
{Pineau des Forets}, G., {Roueff}, E., \& {Flower}, D.~R. 1992, \mnras, 258,
  45P

\bibitem[{{Plasson} {et~al.}(2011){Plasson}, {Brandenburg}, {Jullien}, \&
  {Bersini}}]{2011JPCA..115.8073P}
{Plasson}, R., {Brandenburg}, A., {Jullien}, L., \& {Bersini}, H. 2011, J.
  Phys. Chem., 115, 8073

\bibitem[{{Prasad} \& {Tarafdar}(1983)}]{1983ApJ...267..603P}
{Prasad}, S.~S. \& {Tarafdar}, S.~P. 1983, \apj, 267, 603

\bibitem[{{Rednyk} {et~al.}(2019){Rednyk}, {Rou{\v{c}}ka}, {Kovalenko}, {Tran},
  {Dohnal}, {Pla{\v{s}}il}, \& {Glos{\'\i}k}}]{2019A&A...625A..74R}
{Rednyk}, S., {Rou{\v{c}}ka}, {\v{S}}., {Kovalenko}, A., {et~al.} 2019, \aap,
  625, A74

\bibitem[{{Roueff} \& {Le Bourlot}(2020)}]{2020A&A...643A.121R}
{Roueff}, E. \& {Le Bourlot}, J. 2020, \aap, 643, A121

\bibitem[{{Sagu\'es} \& {Epstein}(2003)}]{2003DaltTra...7..1201S}
{Sagu\'es}, F. \& {Epstein}, I.~R. 2003, Dalton Trans., 1201

\bibitem[{{Schlemmer} {et~al.}(2005){Schlemmer}, {Asvany}, \&
  {Giesen}}]{2005PCCP....7.1592S}
{Schlemmer}, S., {Asvany}, O., \& {Giesen}, T. 2005, Phys. Chem. Chem. Phys.,
  7, 1592

\bibitem[{{Shalabiea} \& {Greenberg}(1995)}]{1995A&A...296..779S}
{Shalabiea}, O.~M. \& {Greenberg}, J.~M. 1995, \aap, 296, 779

\bibitem[{{Tyson}(1975)}]{1975JChPh..62.1010T}
{Tyson}, J.~J. 1975, \jcp, 62, 1010

\bibitem[{{Viti} {et~al.}(2001){Viti}, {Roueff}, {Hartquist}, {Pineau des
  For{\^e}ts}, \& {Williams}}]{2001A&A...370..557V}
{Viti}, S., {Roueff}, E., {Hartquist}, T.~W., {Pineau des For{\^e}ts}, G., \&
  {Williams}, D.~A. 2001, \aap, 370, 557

\bibitem[{{Wakelam} {et~al.}(2006){Wakelam}, {Herbst}, {Selsis}, \&
  {Massacrier}}]{2006A&A...459..813W}
{Wakelam}, V., {Herbst}, E., {Selsis}, F., \& {Massacrier}, G. 2006, \aap, 459,
  813

\bibitem[{{Wirstr{\"o}m} {et~al.}(2016){Wirstr{\"o}m}, {Adande}, {Milam},
  {Charnley}, \& {Cordiner}}]{2016IAUFM..29A.271W}
{Wirstr{\"o}m}, E.~S., {Adande}, G., {Milam}, S.~N., {Charnley}, S.~B., \&
  {Cordiner}, M.~A. 2016, IAU Focus Meeting, 29A, 271

\bibitem[{{Wirstr{\"o}m} \& {Charnley}(2018)}]{2018MNRAS.474.3720W}
{Wirstr{\"o}m}, E.~S. \& {Charnley}, S.~B. 2018, \mnras, 474, 3720

\bibitem[{{Wirstr{\"o}m} {et~al.}(2012){Wirstr{\"o}m}, {Charnley}, {Cordiner},
  \& {Milam}}]{2012ApJ...757L..11W}
{Wirstr{\"o}m}, E.~S., {Charnley}, S.~B., {Cordiner}, M.~A., \& {Milam}, S.~N.
  2012, \apjl, 757, L11

\end{thebibliography}

\end{document}